\documentclass[11pt]{amsart} 
\usepackage{amscd,amssymb,amsxtra}
\usepackage[mathscr]{eucal}% \usepackage{mathrsfs}  % \mathscr{A}
\usepackage{comment}
\usepackage{color}
\usepackage{enumerate}
\makeatletter
\let\@secnumfont\bfseries
\def\section{\@startsection{section}{1}%
  \z@{4\linespacing\@plus\linespacing}{\linespacing}%
  {\bfseries\centering}}
\def\introsection{\@startsection{section}{1}%
  \z@{3\linespacing\@plus\linespacing}{\linespacing}%
  {\bfseries\centering}}
\def\subsection{\@startsection{subsection}{2}%
   \z@{1.25\linespacing\@plus.7\linespacing}{.5\linespacing}%
   {\normalfont\bfseries}}
\def\subsectionsinline{\def\subsection{\@startsection{subsection}{2}%
  \z@{1\linespacing\@plus.7\linespacing}{-.5em}%
  {\normalfont\bfseries}}}

\makeatother

\theoremstyle{definition}
\newtheorem{definition}[equation]{Definition}
\newtheorem{example}[equation]{Example}

\newtheorem*{definition*}{Definition}
\newtheorem*{example*}{Example}
\newtheorem*{problem*}{Problem}
\newtheorem*{exercise*}{Exercise}
\newtheorem*{question*}{Question}
\newtheorem*{construction*}{Construction}

\theoremstyle{remark}

\newtheorem{remark}[equation]{Remark}
\newtheorem{data}[equation]{Data}

\newtheorem*{note*}{Note}
\newtheorem*{notation*}{Notation}
\newtheorem*{remark*}{Remark}
\newtheorem*{data*}{Data}

\theoremstyle{plain}

\newtheorem{claim}[equation]{Claim}

\newtheorem*{theorem*}{Theorem}
\newtheorem*{corollary*}{Corollary}
\newtheorem*{lemma*}{Lemma}
\newtheorem*{proposition*}{Proposition}
\newtheorem*{conjecture*}{Conjecture}
\newtheorem*{claim*}{Claim}
\newtheorem*{proposal*}{Proposal}
\newtheorem*{conclusion*}{Conclusion}
\newtheorem*{hypothesis*}{Hypothesis}

\numberwithin{equation}{section}

\definecolor{refkey}{rgb}{0,.6,.4}

\renewcommand{\:}{\colon}
\DeclareMathOperator{\Aut}{Aut}
\newcommand{\CC}{{\mathbb C}}

\DeclareMathOperator{\Hom}{Hom}
\DeclareMathOperator{\id}{id}
\DeclareMathOperator{\Map}{Map}

\DeclareMathOperator{\pt}{pt}
\newcommand{\QQ}{{\mathbb Q}}

\newcommand{\RR}{{\mathbb R}}
\newcommand{\TT}{\mathbb T}
\DeclareMathOperator{\Spin}{Spin}

\newcommand{\ZZ}{{\mathbb Z}}

\newcommand{\chiup}{\raise.5ex\hbox{$\chi$}}
\newcommand{\cir}{S^1}

\newcommand{\dbar}{{\bar\partial}}

\newcommand{\mstrut}{^{\vphantom{1*\prime y\vee M}}}

\newcommand{\res}[1]{\negmedspace\bigm|\mstrut_{#1}}
\newcommand{\temsquare}{\raise3.5pt\hbox{\boxed{ }}}

\usepackage[all,2cell,dvips]{xy}\renewcommand{\cir}{\ensuremath{S^1}}
\usepackage{graphicx}
\usepackage[colorlinks,citecolor=refkey]{hyperref}

\definecolor{refkey}{rgb}{0,.8,.2}\definecolor{labelkey}{rgb}{1,0,0} 

\theoremstyle{definition}
\newtheorem{expectation}[equation]{Expectation}

\DeclareMathOperator{\Det}{Det}
\DeclareMathOperator{\Vect}{Vect}
\newcommand{\BNG}{B\mstrut _{\nabla }G}
\newcommand{\BNbG}{B\mstrut _{\nabla }\bG}
\newcommand{\Bp}[1]{B\pi (#1)}
\newcommand{\Btp}[1]{B^2\pi (#1)}
\newcommand{\Gz}{\Gamma _{\mathfrak{z}}}
\newcommand{\Hpd}[2]{H^{#1}(#2;\pi \dual)}
\newcommand{\Hp}[2]{H^{#1}(#2;\pi )}
\newcommand{\TX}{Theory~$\sX$}
\newcommand{\Vi}{\Vect_{\textnormal{top}}}
\newcommand{\Xg}{\sX_{\fg}}
\newcommand{\ag}{\alpha_{\fg}}
\newcommand{\bG}{\overline{G}}
\newcommand{\bM}{\overline{M}}
\newcommand{\bP}{\overline{P}}
\newcommand{\bX}{\widetilde{X}}
\newcommand{\bY}{\widetilde{Y}}
\newcommand{\bT}{\overline{\Theta}}
\newcommand{\bo}{\mathbf{1}}
\newcommand{\dual}{^\vee}
\newcommand{\fg}{\mathfrak{g}}
\newcommand{\fh}{\mathfrak{h}}
\newcommand{\form}{\langle -,-  \rangle}
\newcommand{\gpd}{/\!/} 

\newcommand{\rAut}{\Aut_{\textnormal{rel}}}
\newcommand{\sG}{\mathscr{G}}
\newcommand{\sX}{\mathscr{X}}
\newcommand{\sF}{\mathcal{F}}
\newcommand{\Man}{\mathbf{Man}}
\newcommand{\sSet}{\Set_\Delta }
\newcommand{\Set}{\mathbf{Set}}
\newcommand{\op}{^{\textnormal op}}
\newcommand{\tF}{\widetilde{F}}
\newcommand{\ta}{\tau \mstrut _{\le n}\alpha }

\begin{document}

\abovedisplayskip18pt plus4.5pt minus9pt
\belowdisplayskip \abovedisplayskip
\abovedisplayshortskip0pt plus4.5pt
\belowdisplayshortskip10.5pt plus4.5pt minus6pt
\baselineskip=15 truept
\marginparwidth=55pt

\renewcommand{\labelenumi}{\textnormal{(\roman{enumi})}}

%**end of header

% lasteq@ 38
% lastsec@  6
% lastthm@ 16
% lastfig@  4

%$$\boxed{\boxed{\text{PRELIMINARY VERSION}}}$$\par\vskip 2pc % omit in final

 \title[Relative Quantum Field Theory]{Relative Quantum Field Theory} %% replace \today with short title in final version
 \author[D. S. Freed]{Daniel S.~Freed}
 \thanks{The work of D.S.F. is supported by the National Science Foundation
under grant DMS-1160461}
 \address{The University of Texas at Austin \\ Mathematics Department RLM
8.100 \\ 2515 Speedway Stop C1200\\ Austin, TX 78712-1202}
 \email{dafr@math.utexas.edu}

 \author[C. Teleman]{Constantin Teleman} 
  \thanks{The work of C.T. is supported by NSF grant DMS-1160461}
 \address{Department of Mathematics \\ University of California \\ 970 Evans
Hall \#3840 \\ Berkeley, CA 94720-3840}  
 \email{teleman@math.berkeley.edu}

% \dedicatory{}
 \date{January 14, 2014}
 \begin{abstract} 
 We highlight the general notion of a {\it relative quantum field theory\/},
which occurs in several contexts.  One is in gauge theory based on a compact
Lie algebra, rather than a compact Lie group.  This is relevant to the
maximal superconformal theory in six dimensions.
 \end{abstract}
\maketitle

%\pagestyle{myheadings}   % omit in final
%\markboth{PRELIMINARY VERSION (\today)}{PRELIMINARY VERSION (\today)}  % omit

   \section{Introduction}\label{sec:1}
% lastsubsec@000

The $(0,2)$-superconformal field theory in six dimensions, which we term \TX\
for brevity, was discovered as a limit of superstring theories~\cite{W1,S}.
It is thought not to have a lagrangian description, so is difficult to access
directly, yet some expectations can be deduced from the string theory
description~\cite{W2,GMN}.  Two features are particularly relevant: (i)~it is
not an ordinary quantum field theory, and (ii)~the theory depends on a Lie
algebra, not on a Lie group.  A puzzle, emphasized by Greg Moore, is that the
dimensional reduction of \TX\ to five dimensions is usually understood to be
an ordinary quantum field theory---contrary to~(i)---and it is a
supersymmetric gauge theory so depends on a particular choice of Lie
group---contrary to~(ii).  In this paper we spell out the modified notion
indicated in~(i), which we call a \emph{relative quantum field theory}, and
use it to resolve this puzzle about \TX\ by pointing out that the dimensional
reduction is also a relative theory.  Relative gauge theories are not
particular to dimension five.  In fact, the possibility of studying
four-dimensional gauge theory as a relative theory was exploited
in~\cite{VW} and~\cite{W3}.  
 
A relative quantum field theory~$F$ is related to an \emph{extended} quantum
field theory~$\alpha $ in one higher dimension.  On a compact (Euclidean)
spacetime the partition function of~$F$ is a vector in the quantum Hilbert
space of~$\alpha $, and on a compact space the quantum Hilbert space of~$F$
is also related to the value of~$\alpha $, which is a category.  An
\emph{anomalous} quantum field theory~$F$ may be viewed as a relative
theory.\footnote{That point of view may not always be useful; we give
examples in~\S2.}  The anomaly~$\alpha $ is an \emph{invertible} quantum
field theory: each of its quantum Hilbert spaces is one-dimensional and the
partition functions are nonzero.  The invertibility of~$\alpha $ implies that
the partition functions of~$F$ are defined as numbers only up to a scalar;
the field theory~$\alpha $ controls the indeterminacy.  Another well-known
example is the two-dimensional chiral Wess-Zumino-Witten conformal field
theory, which is a theory relative to three-dimensional topological
Chern-Simons theory.  In these examples, as well as the ones in this paper,
the higher dimensional theory~$\alpha $ obeys strong finiteness conditions,
though the definition does not require that.  For example, the quantum
Hilbert spaces are finite dimensional.  In addition, in our main examples
here $\alpha $~is a \emph{topological} field theory.  (That is not always
true---for example, anomaly theories are generally not topological.)
 
In~\S\S\ref{sec:3}--\ref{sec:5} we study three examples of relative
theories.  The first two are constructed by quantization of a classical
model, whose fields form a fibration 
  \begin{equation}\label{eq:37}
     p\:\sF\longrightarrow \sF'' 
  \end{equation}
Relative fields are the fibers of~$p$; the fiber~$\sF'$ over the basepoint
of~$\sF''$ is special.  In our examples the quotient $\sF''$~is finite in the
sense that the path integral over the field~$\sF''$ reduces to a finite
sum.\footnote{We can make the relative theory completely rigorous if $\sF$~is
also finite in this sense.}  The third example is \TX, for which classical
fields are only a heuristic unless the theory is noninteracting.  The
theory~$\alpha $ in each of our three examples involves a finite group~$\pi
$.  In~\S\ref{sec:3} we study a relative $\sigma $-model, for which $\pi $~is
an \emph{arbitrary} finite group.  There are relative gauge theories based on
finite covers $G\to\bG$ of compact connected Lie groups with covering group
an \emph{abelian} group~$\pi $.  These are discussed in~\S\ref{sec:4}.
In~\S\ref{sec:5} we turn to \TX, and $\pi $~is restricted to be a
\emph{Pontrjagin self-dual} finite abelian group.  The data which
defines~\TX\ is usually taken to be a compact simple real Lie algebra of
type~A, D, or~E.  In Data~\ref{thm:7} we generalize to include noninteracting
theories and many other examples.  Appendix~\ref{sec:6} posits a brief
definition of a field; it is useful for the discussions in the body of the
paper.
 
We begin in~\S\ref{sec:2} with a general discussion of a relative quantum
field theory.  This notion has appeared elsewhere in various guises.  One of
the first is Segal's discussion~\cite[\S5]{S2} of a ``weakly''
two-conformal field theory with associated ``modular functor'';
the modular functor is part of topological Chern-Simons theory.  We already
mentioned Witten's description~\cite{W2} of \TX.  One can view a relative
theory~$F$ as a \emph{boundary theory} for the higher dimensional
theory~$\alpha $, in which case the notion is ubiquitous; in a topological
context it is embedded in Kapustin's discussion~\cite{K}.  It is also the
framework in which Kevin Walker~\cite{Wa} describes Chern-Simons theory, and
it is a very special case of Lurie's notion~\cite{L} of a topological field
theory defined on manifolds with singularities.
 
We thank David Ben-Zvi, Jacques Distler, Greg Moore, and Andy Neitzke for
discussions, and we also thank Edward Witten for comments on the first draft.
We express our gratitude to the Aspen Center for Physics for hospitality.

   \section{Relative quantum theories}\label{sec:2}
% lastsubsec@000

\begin{figure}[ht]
  \centering
  \includegraphics[scale=.8]{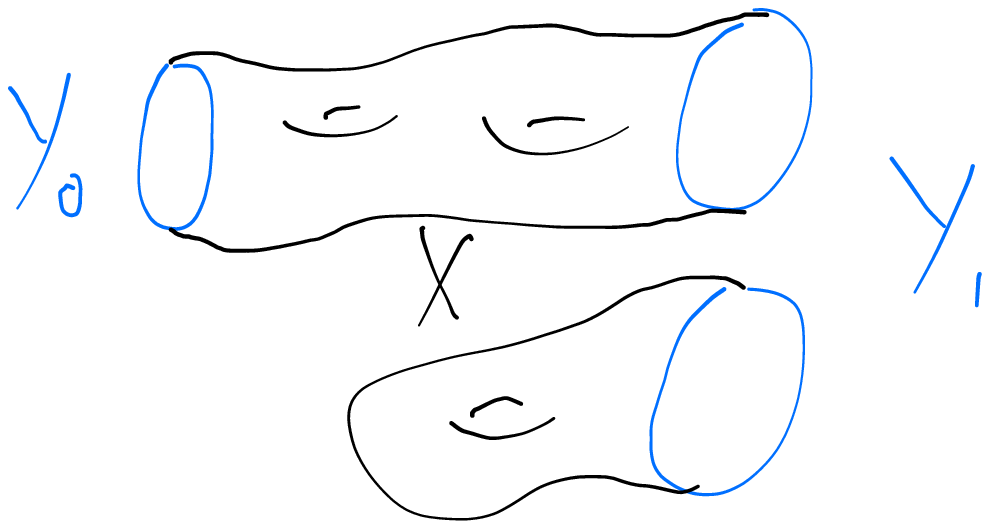}
  \caption{A geometric bordism $X\:Y_0\to Y_1$}\label{fig:1}
  \end{figure}

An $m$-dimensional quantum field theory (QFT)~$f$ assigns to an
$m$-dimensional manifold~$X$ a partition function $f(X)\in \CC$ and to an
$(m-1)$-dimensional manifold~$Y$ a quantum Hilbert space~$f(Y)$.  We have in
mind a theory defined on Riemannian manifolds---so Wick rotated from a theory
on Minkowski spacetime---though it may be a conformal theory or a topological
theory.  The manifolds~$X,Y$ may also carry topological structure, such as an
orientation, spin structure, or framing.  Finally, all manifolds are assumed
compact to avoid convergence issues, and above $X,Y$~do not have boundary.
The theory is also defined for compact $m$-manifolds~$X$ with boundary
$\partial X=Y_0\amalg Y_1$ expressed as a disjoint union of
closed\footnote{For manifolds `closed'=`compact without boundary'.}
manifolds, and viewed as a map $X\:Y_0\to Y_1$ (see Figure~\ref{fig:1}).
Then $f(X)\:f(Y_0)\to f(Y_1)$ is a linear map on the quantum Hilbert spaces.
For example, if $X$~is a closed $m$-manifold with ~$x_1,\dots ,x_k\in X$,
define~$X_\epsilon $ as $X$~with open balls of radius~$\epsilon $ about
each~$x_i$ omitted, and view the boundary spheres as incoming.  In the
limit~$\epsilon \to0$ the theory gives a map $V\times \cdots\times V\to \CC$
is the correlation function on the space~$V$ of operators attached to a
point.  The modern mathematical take is that a quantum field theory~ $f$ is a
homomorphism from a geometric bordism category to a category~$\Vi$ of
topological vector spaces; the Hilbert structure emerges under special
conditions.\footnote{Therefore, we use `quantum topological vector space' in
place of the usual `quantum Hilbert space'.}  We do not give a precise
formulation here.\footnote{See~\cite{S1} for a recent discussion of geometric
axioms for quantum field theory.}  An \emph{extended} quantum field
theory~$f$ also assigns values to closed $(m-2)$-manifolds~$Z$ and
$(m-1)$-manifolds with boundary: $f(Z)$~is a linear category
whose hom-sets are topological vector spaces.  The extended theories in this
paper (denoted~`$\alpha $') are finite dimensional---both the quantum
spaces~$f(Y)$ and the hom-sets in the linear category~$f(Z)$ are finite
dimensional vector spaces---whereas the relative theories (denoted~`$F$') are
typically infinite dimensional.

A theory~$f$ is \emph{invertible} if $f(X)\not= 0$ for all $m$-manifolds, the
vector space~$f(Y)$ is one-dimensional for all $(m-1)$-manifolds, and the
linear category~$f(Z)$ is similarly invertible for all $(m-2)$-manifolds: it
is a free $\Vi$-module of rank one.  The trivial theory~$\bo$ is the constant
invertible theory with values $\bo(X)=1$, $\bo(Y)=\CC$, and $\bo(Z)=\Vi$.

Here is a concise formal definition of a relative
QFT.

  \begin{definition}[]\label{thm:1}
 Fix an integer~$n\ge0$ and let $\alpha $~be an extended $(n+1)$-dimensional
quantum field theory.  A \emph{quantum field theory~$F$ relative to~$\alpha
$} is a homomorphism
  \begin{equation}\label{eq:1}
     F\:\bo\longrightarrow \ta.
  \end{equation} 
  \end{definition}

\noindent
 The relative theory ignores the partition functions and correlation
functions of~$\alpha $ on $(n+1)$-manifolds and consider only the
truncation~$\ta$ to a theory of $n$- and $(n-1)$-manifolds.  In some cases
the relative theory is more naturally a map in the other direction:
  \begin{equation}\label{eq:26}
     \tF\:\ta\longrightarrow \bo. 
  \end{equation}
We now spell out what data is contained in a homomorphism~$\tF$; the story is
similar for~\eqref{eq:1}.

Let $X$~be a closed $n$-manifold.  Then $\alpha (X)\in \Vi$ is a topological
vector space and $\bo(X)=\CC$.  As mentioned above, in all examples
considered here $\alpha (X)$~is finite dimensional.  The relative theory~$\tF$
assigns to~$X$ a linear functional $\tF(X)\:\alpha (X)\to\CC$.  So for each
vector~$\xi \in \alpha (X)$ there is a partition function 
  \begin{equation}\label{eq:4}
     \tF(X;\xi )\in \CC. 
  \end{equation}

Let $Y$~be a closed $(n-1)$-manifold.  Then $\alpha (Y)$~is a linear category
and $\bo(Y)=\Vi$.  The relative theory assigns to~$Y$ a homomorphism
$\tF(Y)\:\alpha (Y)\to\Vi$.  Thus for each object~$\mu $ in the
category~$\alpha (Y)$ there is a quantum topological vector space
  \begin{equation}\label{eq:5}
     \tF(Y;\mu )\in \Vi. 
  \end{equation}

The  situation for  a  compact $n$-manifold  $X\:Y_0\to  Y_1$ is  a bit  more
complicated.    The   categories~$\alpha    (Y_0),\alpha   (Y_1)$   and   the
maps~$\tF(Y_0),\tF(Y_1)$ fit into the commutative diagram
  \begin{equation}\label{eq:2}
     \xymatrix@C+24pt@R-12pt{\alpha (Y_0)\ar[dd]_{\alpha (X)}\ar[dr]^{\tF(Y_0)}\\ 
     &\Vi\\ 
     \alpha (Y_1)\ar[ur]_{\tF(Y_1)}} 
  \end{equation}
The relative theory assigns to~$X$ a homomorphism 
  \begin{equation}\label{eq:3}
     \tF(X)\: \tF(Y_1)\circ \alpha (X)\longrightarrow \tF(Y_0). 
  \end{equation}

As a check on the definition, a QFT~$f$ relative to the trivial theory~$\bo$
is an \emph{absolute} $n$-dimensional QFT,\footnote{The arrow in~\eqref{eq:3}
is opposite to what we expect if~$\alpha =\bo$, but $\tF(X^\vee)^\vee$ does
point in the right direction, where `${}^\vee$'~denotes the dual bordism and
the dual linear map.} where we use `absolute' to describe a usual quantum
field theory as opposed to a relative one.
 
If $\alpha $~is an \emph{invertible} $(n+1)$-dimensional theory, and $\tF$~is
a theory relative to~$\alpha $, then we say \emph{$\tF$~is anomalous with
anomaly~$\alpha $}.  In this case $\alpha (X)$~is one-dimensional and there
is a single partition function~\eqref{eq:4} determined up to a scalar
controlled by~$\alpha (X)$.  If $\xi \in \alpha (X)$~is nonzero, then any
other vector has the form~$\xi '=\lambda \xi $ for some~$\lambda \in \CC$,
and then $\tF(X;\xi ')=\lambda \tF(X;\xi )$.  If $\alpha $~is a unitary
theory, then we can choose~$\xi ,\xi '$ to have unit norm, in which case the
partition function is determined up to a phase.  Similarly, the quantum
topological vector space~\eqref{eq:5} is determined up to a vector space
controlled by~$\alpha (Y)$.  In the unitary case we can restrict to ``unit
norm'' objects $\mu \in \alpha (Y)$, which comprise a gerbe, and if $\mu
'=L\otimes \mu $ for a complex line~$L$ then $\tF(Y;\mu ')=L\otimes \tF(Y;\mu
)$.  In particular, the underlying projective space of~$\tF(Y;\mu )$ is
independent of~$\mu $.  This is the standard picture of an anomalous theory.
 
A concrete example is provided by the conformal anomaly of an
$n=2$~dimensional conformal field theory~$f$.  The theory~$f$ is an ordinary
theory of oriented \emph{Riemannian} manifolds of dimension~1 and~2, or more
precisely an anomalous theory of oriented \emph{conformal} manifolds of the
same dimensions.  The conformal anomaly theory~$\alpha $ is 3-dimensional,
but we only consider its truncation\footnote{On a closed 3-manifold the
partition function is the exponential of a multiple of an $\eta $-invariant.
It does not play a role in the 2-dimensional truncation} to manifolds of
dimension~$\le2$.  On an oriented conformal surface~$X$,
  \begin{equation}\label{eq:6}
     \alpha (X) = (\Det _X)^{\otimes c_L}\otimes (\overline{\Det _X})^{\otimes 
     c_R} ,
  \end{equation}
where $\Det _X$ is the determinant line of the $\dbar$-operator determined by
the conformal structure and orientation, and $c_L,c_R\in \RR$ differ by an
integer; the latter condition implies that the partition function is
well-defined without further tangential structure on~$X$.  See~\cite[\S4]{S2}
for more discussion.

Here are two examples of relative field theories with $\alpha $~invertible.
In the first case $F_1$~is a 4-dimensional gauge theory with chiral fermions,
so the fermionic path integral has an anomaly.  The partition function
depends on~$(X,g,A)$, where $X$~is a closed 4-manifold, $g$~a Riemannian
metric, and $A$~a connection (gauge field), and it takes values in a
determinant or pfaffian line ~$\alpha _1(X,g,A)$.  One cannot treat~$A$ as a
quantum field due to the anomaly, but the theory makes good sense with $A$~as
a background field.  In the second case $F_2$~is the two-dimensional gauged
WZW~model~\cite{W4}.  The partition function depends on~$(X,g,A)$ for a
closed oriented 2-manifold~$X$ with metric~$g$ and connection~$A$, and it
takes values in the Chern-Simons line~$\alpha _2(X,A)$.  In both cases the
invertible anomaly theory~$\alpha $ can be regarded as \emph{classical},
rather than quantum.  Indeed, a classical field theory is an example of an
invertible field theory, and the formulation in terms of geometric bordism
categories does not distinguish classical from quantum.  One can quantize the
pair~$(\alpha _2,F _2)$.  The quantization\footnote{There is a well-known
framing anomaly, and quantum Chern-Simons is defined on a bordism category
which includes framings, as described in the next paragraph.} of classical
Chern-Simons theory~$\alpha _2$ is quantum Chern-Simons theory~$\hat\alpha
_2$.  The function~$F_2(X,g)(A)$ of~$A$, which is a section of the
Chern-Simons line bundle $\alpha _2(X)$~over the space of gauge fields~$A$,
is a vector~$F_2(X,g)$ in the vector space~$\hat\alpha _2(X)$ of quantum
Chern-Simons theory~$\hat\alpha _2$.  (There is a polarization
condition---holomorphy---which is part of the quantization~\cite[\S2.1]{W4}.)
The pair of quantum theories~$(\hat\alpha _2,\widehat F_2)$~is mentioned in
the next paragraph, where it is denoted~$(\chi ,\sF)$.  Note that $\hat\alpha
_2$~is almost never invertible.
 
Chiral, or holomorphic, conformal field theories in $n=2$~dimensions are a
motivating example for Definition~\ref{thm:1}.  We use Riemannian manifolds
to avoid the conformal anomaly.  The manifolds must be oriented and carry an
additional topological structure which has several alternative descriptions:
a 2-framing~\cite{A}, a $p_1$-structure~\cite{BHMV}, or a rigging~\cite{S2}.
For concreteness let us consider the chiral WZW model~$\tF$ based on a
compact Lie group~$G$ and a level~$k$.  Then there is an associated
3-dimensional \emph{topological} field theory~$\chi $, the Chern-Simons
theory associated to~$(G,k)$.  It assigns a vector space~$\chi (X)$ to an
oriented rigged surface.  In the chiral WZW~model, $\chi (X)$~is interpreted
as the space of conformal blocks and there is a partition
function~\eqref{eq:4} for each conformal block.  The linear category~$\chi
(S^1)$ attached to the circle~$Y=\cir$ is a modular tensor category, which in
many cases has a combinatorial description.  For example, if $G$~is
1-connected it can be described in terms of a quantum group.  But the
description of~$\chi (\cir)$ as the category of positive energy
representations of the loop group of~$G$ at level~$k$ is more adapted to the
WZW~model: the topological vector space~$\tF(\cir;\mu )$ \emph{is} the
underlying space of the representation~$\mu \in \chi (\cir)$.  The
truncation~$\tau \mstrut _{\le2}\chi $ is called a \emph{modular functor}
in~\cite[\S5]{S2} and the relative theory~$\tF$ is called a \emph{weakly
conformal field theory}.  (It is formulated on conformal surfaces, rather
than Riemannian surfaces, and there is an additional holomorphy condition.)

The topological Chern-Simons theory determines a modular tensor category 
$A = \chi(S^1)$ (up to equivalence of categories). The WZW model specifies a 
weak braided tensor functor $A\to\Vi$.

  \begin{figure}[ht]
  \centering
  \ \hskip-90pt\includegraphics[scale=.8]{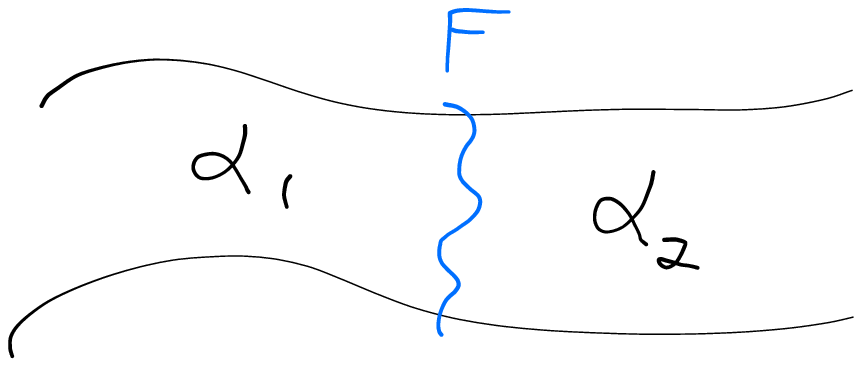}
  \caption{A domain wall}\label{fig:2}
  \end{figure}

We conclude by briefly indicating the relationship of a relative quantum
field theory to other variations of standard quantum field theories.  First,
if $\alpha _1,\alpha _2$~are $(n+1)$-dimensional theories, then a
\emph{domain wall} is an $n$-dimensional theory which lives on a codimension
one submanifold of an $(n+1)$-manifold on which $\alpha _1$~ and $\alpha
_2$~are defined, as depicted in Figure~\ref{fig:2}.  See~\cite{K} for a
discussion in the context of extended topological theories.  To match
Definition~\ref{thm:1} put $\alpha _1=\alpha $ and $\alpha _2=\bo$.  Ignoring
the trivial theory we obtain from Figure~\ref{fig:2} an $(n+1)$-manifold with
boundary.  The theory~$\alpha $ lives in the bulk and the theory~$\tF$ on the
boundary.  So a relative theory $\tF$~may be regarded as a \emph{boundary
theory} for~$\alpha $.
 
Finally, in the context of topological field theories there is a vast
generalization based on bordism categories of manifolds with
singularities~\cite[\S4.3]{L}.  The appropriate ``singular'' manifold for a
relative theory is the cone on a point.  The cobordism hypothesis asserts
that a fully extended\footnote{A fully extended $n$-dimensional theory is
defined for all manifolds of dimension~$\le n$.  The values increase in
category number as the dimension decreases.} topological theory is determined
by its value on a point, and the extension to manifolds with singularities
implies the same for a fully extended relative theory in which both $\alpha
$~and $\tF$~are topological.

In rest of the paper $\alpha $~is a topological theory, the relative
theory~$F$ more naturally maps in the direction~\eqref{eq:1}, and $F$~is not
topological.

   \section{Warmup: relative $\sigma $-models}\label{sec:3}
% lastsubsec@000

In this section we interpret the familiar example of a $\sigma $-model as a
relative quantum field theory.  We show how formal quantization of classical
fields leads to the relative QFT structure.  The classical theory is defined
in any dimension~$n$.

  \begin{data}[]\label{thm:2}
 \ 

  \begin{enumerate}
 \item $\pi $~a finite group 

 \item $M$~ a smooth manifold with a free left $\pi $-action and quotient~$\bM$.
  \end{enumerate}
  \end{data}

\noindent
 We study the $\sigma $-model into~$\bM$, or equivalently the gauged $\sigma
$-model into~$M$.
 
Let $\Bp X$ denote the collection of principal $\pi $-bundles over~$X$.
Recall that a principal $\pi $-bundle, or Galois covering space with Galois
group~$\pi $, is a covering space $P\to X$ and a free $\pi $-action on~$P$
such that $P\to X$ is a quotient map for the $\pi $-action.  The collection
of $\pi $-bundles over~$X$ is a \emph{groupoid}, not a space.  This is to
account for symmetries of fields: a symmetry 
  \begin{equation}\label{eq:27}
     \varphi \:(P\to X)\longrightarrow (P'\to X) 
  \end{equation}
is a diffeomorphism $\varphi \:P\to P'$ which commutes with the $\pi $-action
and covers~$\id_X$.  The automorphism group of $(P\to X)$ is the group of
gauge transformations.  The path integral over~$\Bp X$ is an integral over
the equivalence classes of $\pi $-bundles.  Canonical quantization over~$\Bp
Y$ for an $(n-1)$-manifold~$Y$ also remembers the gauge symmetry (Gauss law).

The space of fields of the $\sigma $-model into~$\bM$ on a manifold~$X$ is
the space~$\Map(X,\bM)$ of smooth maps $f\:X\to \bM$.  A $\sigma $-model
field induces a gauge field: define
  \begin{equation}\label{eq:7}
     \begin{aligned} p\:\Map(X,\bM) &\longrightarrow \Bp X \\ f&\longmapsto
      f^*(M\to\bM)\end{aligned} 
  \end{equation}
That is, to a map $f\:X\to\bM$ the map~$p$ assigns the pullback of the $\pi
$-bundle $M\to \bM$.  This pullback is the obstruction to lifting~$f$ to a
map $X\to M$: a lift is precisely a trivialization of the pullback $f^*(M\to
\bM)$.  The map~$p$ need not be surjective.  For example, in the extreme case
$M=\bM\times \pi $ the image of~$p$ contains only the trivial $\pi $-bundle
over~$X$.

Relative fields are the fibers of the map~$p$ in~\eqref{eq:7}.

  \begin{definition}[]\label{thm:3}
 Fix $(P\to X)\in \Bp X$.  A \emph{relative field} over $(P\to X)$ is a
pair~$(f,\theta )$ consisting of a smooth map $f\:X\to\bM$ and an isomorphism
$\theta \:(P\to X)\longrightarrow f^*(M\to\bM)$ of $\pi $-bundles over~$X$.
  \end{definition}

\noindent
 Equivalently, a relative field~$(f,\theta )$ is a $\pi $-equivariant map
$P\to M$.  In particular, relative fields are rigid---there are no
automorphisms---so form a space, not a groupoid.  Notice that the fiber
of~$p$ over the trivial bundle $(X\times \pi \to X)$ is the mapping
space~$\Map(X,M)$.

We now indicate a pair~$(\alpha ,F)$ of theories in which $\alpha $~is
topological, defined on all manifolds, and $F$~is a Riemannian theory.  We
use our knowledge of the $\sigma $-model~$F$ to predict the structure
of~$\alpha $.  To define the classical $\sigma $-model, we assume $\bM$~
carries geometric data---a metric, $B$-field, etc.---which is lifted to $\pi
$-invariant geometric data on~$M$.  Formally, the \emph{relative path
integral}~$F(X)$ on a closed $n$-manifold~$X$ is an integral over relative
fields, so a function $F(X)\:\Bp X\to\CC$.  The formal structure implies that
$F(X)$~is invariant under symmetries in~$\Bp X$, so passes to a function on
equivalence classes.  Let $H^1(X;\pi )$~denote the set of equivalence
classes, which are isomorphism classes of principal $\pi $-bundles over~$X$.
Then
  \begin{equation}\label{eq:8}
     F(X)\:H^1(X;\pi )\longrightarrow \CC. 
  \end{equation}
If $F$~is to be a QFT relative to an $(n+1)$-dimensional theory~$\alpha $ in
the sense that $F\:\bo\to \ta$, then $F(X)\:\CC\to \alpha (X)$.  In other
words, $F(X)$~can be identified as an element of the vector space~$\alpha
(X)$.  This leads to the prediction that $\alpha (X)$~is the free vector
space
  \begin{equation}\label{eq:9}
     \alpha (X)=\CC\left\{ H^1(X;\pi ) \right\} 
  \end{equation}
on the finite set~$H^1(X;\pi )$.

  \begin{figure}[ht]
  \centering
  \includegraphics[scale=.8]{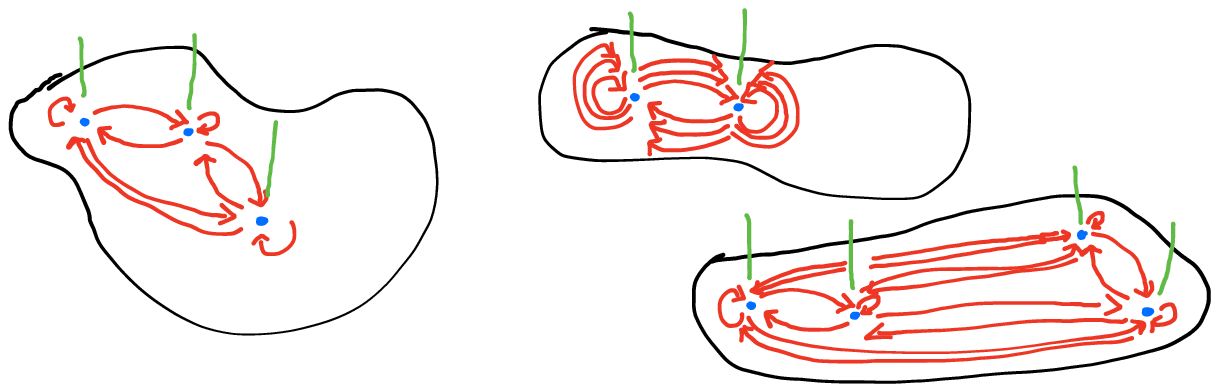}
  \caption{The vector bundle $F(Y)\to\Bp Y$}\label{fig:3}
  \end{figure}

\begin{figure}[ht]
  \centering
  \includegraphics[scale=.8]{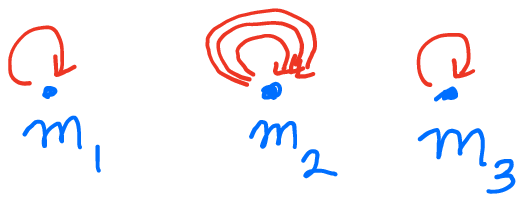}
  \caption{A groupoid equivalent to~$\Bp Y$}\label{fig:4}
  \end{figure}

Now let $Y$~be a closed $(n-1)$-manifold.  The \emph{relative canonical
quantization}~$F(Y)$ is obtained by carrying out canonical quantization on
the fibers of $\Map(Y,\bM)\to\Bp Y$.  Thus $F(Y)\to\Bp Y$ is a vector bundle
(whose fibers are typically infinite dimensional topological vector spaces).
Furthermore, it is an \emph{equivariant} vector bundle: symmetries of fields
in~$\Bp Y$ come with a lift to the vector bundle.  We depict it in
Figure~\ref{fig:3}.  The blue dots represent $\pi $-bundles~$Q\to Y$ and the
red arrows represent isomorphisms of $\pi $-bundles.  The bundles are grouped
into isomorphism classes.  The groupoid~$\Bp Y$ is equivalent\footnote{To
make the equivalence, which is noncanonical, we choose a representative
bundle $Q\to Y$ in each isomorphism class.} to a much simpler groupoid which
has a finite set of objects~$H^1(Y;\pi )$ and in which there are no arrows
between distinct objects; see Figure~\ref{fig:4}.  If $m\in H^1(Y;\pi )$ is
the class of a $\pi $-bundle $Q\to Y$, then the automorphism group of~$m$ is
the group~$\Aut(Q\to Y)$ of gauge transformations of~$Q\to Y$.  The
equivariant bundle
  \begin{equation}\label{eq:14}
     F(Y)\longrightarrow \Bp Y 
  \end{equation}
therefore decomposes into topological vector spaces indexed by pairs~$(m,e)$ in which
$m\in H^1(Y;\pi )$ and $e$~is an irreducible complex representation of the
automorphism group of~$m$.  Now $F(Y)\:\bo(Y)\to\alpha(Y)$~ may be identified
with an object of the linear category~$\alpha (Y)$.  This leads to the
prediction that $\alpha (Y)$~is the free $\Vi$-module with basis
pairs~$(m,e)$.

The special case~$n=1$ is most familiar.  (It has the added advantage that
the quantum theory makes sense.)  If $X=\cir$ in~\eqref{eq:8}, then
$H^1(\cir;\pi )$~is the set of conjugacy classes in~$\pi $.  If $Y=\pt$ is a
single point, then $H^1(\pt;\pi )$ has a single element which represents the
trivial bundle and has automorphism group~$\pi $.  We identify~$\alpha (\pt)$
as the category of representations of~$\pi $.  The entire field
theory~$\alpha $ is familiar: it is the finite 2-dimensional gauge theory
with gauge group~$\pi $.  It may be defined On \emph{all} manifolds of
dimension~$\le2$ by a \emph{finite path integral}~\cite{F,FHLT} as we now
briefly review.
 
The lagrangian of the theory vanishes, so the exponentiated action function
is constant.  On a compact 2-manifold~$W$ the constant is~1, whence the path
integral is a weighted count of $\pi $-bundles:
  \begin{equation}\label{eq:10}
     \alpha (W) = \sum\limits_{[R\to W]}\frac{1}{\#\Aut(R\to W)}. 
  \end{equation}
The sum is over equivalence classes.  The finite path integral over~$X=\cir$
is a sum on $\pi $-bundles over~$\cir$ of the exponentiated action, which is
the constant function that assigns to each $\pi $-bundle $Q\to \cir$ the
trivial complex line~$\CC$.  The result~$\alpha (\cir)$ is the space of
invariant sections of the trivial equivariant line bundle over~$\Bp\cir$.
Since $\Bp\cir$ is equivalent to the quotient groupoid~$G\gpd G$ of
$G$~acting on itself by conjugation, this is the space of central functions
on~$\pi $.  Finally, the exponentiated action on the codimension
two\footnote{The zero-dimensional manifold~$Y=\pt$ has codimension two in a
two-dimensional theory.}  manifold~$Y=\pt$ has constant value the linear
category~$\Vi$.  In this case the finite path integral returns the
subcategory of ``invariants'' under~$\pi $, which is the category~$\alpha
(\pt)$ of representations of the group~$\pi $, as predicted above.  The
finite gauge theory can be defined in any dimension~$n+1$.  Also, there is a
twisted version defined using a group cocycle, which can be incorporated into
the $\sigma $-model as a topological term in the action.

The picture of~$F$ as a \emph{boundary} theory for~$\alpha $ is manifest in
terms of classical fields.  If $W$~is an $(n+1)$-manifold with boundary, and
$R\to W$ a principal $\pi $-bundle---a field in the bulk theory~$\alpha
$---then the boundary field~$f$ is a $\pi $-equivariant function $P\to M$,
where $P\to\partial W$ is the restriction of $R\to W$ to the boundary.  The
relative theory on an $n$-manifold~$X$ has a bulk field $R\to[0,1]\times X$,
a boundary field~$f$ on~$\{1\}\times X$, and no additional field
 at~$\{0\}\times X$.  (At $\{0\}\times X$ lies the trivial theory~$\bo$, whose
space of classical fields consists of a single point.)
 
We remark that the relative theory can be defined in terms of topological
disorder operators in the $\sigma $-model to~$M$.  For example, to define the
partition function~\eqref{eq:8} represent a class in~$H^1(X;\pi )$ by a $\pi
$-bundle $P\to X$ and endow it with a trivialization away from a normally
oriented codimension one ``defect'' submanifold~$D\subset X$.  The
trivialization determines a locally constant function $j\:D\to\pi $, the jump
across~$D$.  Then a relative field determines a function $X\setminus D\to M$
which obeys the jump~$j$.  Notice that this function is only defined away
from~$D$, so does not contain all of the information of the relative field.
 
We can use the relative theory to recover absolute theories.  Namely, if $\pi
'<\pi $ is a subgroup, then we can recover the $\sigma $-model into~$M/\pi '$
by ``\,integrating over~$B\pi '$\,''.  So for a closed $n$-manifold~$X$ we
define
  \begin{equation}\label{eq:12}
     f_{\pi '}(X) = \sum\limits_{m'\in H^1(X;\pi ')}\frac{1}{\#Z_{\pi
     '}(m')}\;F(X;m'), 
  \end{equation}
where the sum is over equivalence classes of $\pi '$-bundles $P'\to X$ and
$Z_{\pi '}(m')$~is the automorphism group of a representative of the
equivalence class~$m'$.  (While the group~$Z_{\pi '}(m')$ depends on the
choice of representative, its cardinality does not.)  Similarly, the quantum
topological vector space on a closed $(n-1)$-manifold is
  \begin{equation}\label{eq:13}
     f_{\pi '}(Y) = \bigoplus\limits_{m'=[Q'\to Y]\in H^1(Y;\pi
     ')}F(Y;Q\to Y)^{\Aut(Q'\to Y)}. 
  \end{equation}
The sum is over equivalence classes of $\pi '$-bundles with chosen
representatives.  A $\pi '$-bundle $Q'\to Y$ has an associated $\pi $-bundle
$Q\to Y$, and automorphisms of~$Q'\to Y$ induce automorphisms of $Q\to Y$.
The summand in~\eqref{eq:13} is the subspace of invariants of the fiber
of~\eqref{eq:14} at~$Q\to Y$.  The extreme cases~$\pi '=\{e\}$ and $\pi '=\pi
$ give the $\sigma $-models into~$M$ and~$\bM$, respectively.  Clearly the
relative theory encodes more information than the absolute theories. 
 
We use the language of Appendix~\ref{sec:6} to describe the relative
fields.\footnote{So \eqref{eq:15}~is a map of simplicial sheaves (or sheaves
of groupoids or stacks) on the category of smooth manifolds.}  Namely,
\eqref{eq:7}~is a map
  \begin{equation}\label{eq:15}
     p\:\bM\longrightarrow B\pi 
  \end{equation}
with fiber~$M$.  In fact, $p$~is the classifying map of the principal $\pi
$-bundle $ M\to\bM$.  Quantization in the relative theory is integration over
the fibers of~$p$ (on a particular manifold~$X$).  The inclusion $i\:\pi
'\to\pi $ of a subgroup induces a pullback diagram
  \begin{equation}\label{eq:16}
     \xymatrix{M/\pi '\ar[r]^{} \ar[d]_{p'} & \bM\ar[d]^{p} \\ B\pi
     '\ar[r]^{Bi} & B\pi } 
  \end{equation}
On a manifold~$X$ integration over the fibers of~$p'$ followed by integration
over~$B\pi '(X)$ is equivalent to integration over~$M/\pi '$.  This explains
the formulas in the previous paragraph.  In these terms the topological
vector space~$f_{\pi '}(Y)$ of~\eqref{eq:13} is the space of invariant
sections of $\bigl(Bi\bigr)^*\bigl(F(Y)\to \Bp Y \bigr)$.

Our hypothesis in Data~\ref{thm:2} is that $\pi $~acts \emph{freely} on~$M$.
We can generalize to arbitrary $\pi $-actions if we interpret~$\bM=M\gpd\pi $
as the stack quotient, so the $\sigma $-model into~$\bM$ as the gauged
$\sigma $-model on~$M$.

   \section{Relative gauge theories}\label{sec:4}
% lastsubsec@000

The relative theory in this section is an exact analog of the relative
$\sigma $-model of~\S\ref{sec:3} with all of the fields bumped up one
categorical level---there is an extra layer of symmetry.

  \begin{data}[]\label{thm:4}
 A covering homomorphism $G\to\bG$ of compact connected Lie groups with
kernel~$\pi $
  \end{data}

\noindent
 So $\pi $~is a finite central subgroup of~$G$, necessarily abelian.  We
denote the common Lie algebra of~$G$ and~$\bG$ as~$\fg$.  

  \begin{example}[]\label{thm:15}
 If $\fg$~is a real algebra with negative definite Killing form, then there
is a canonically associated compact 1-connected Lie group~$G$.  Let $\pi
\subset G$ be the center and $\bG=G/\pi $ the adjoint group.  This gives a
canonical choice of Data~\ref{thm:4} associated to a compact semisimple Lie
algebra.  The more general data allows for torus factors as well.
Simple representative examples are the covers $SU(2)\to SO(3)$, $U(2)\to
U(2)/\{\pm1\}$, and $\TT\xrightarrow{\textnormal{$\lambda \mapsto\lambda
^2$}}\TT $, each with $\pi $~cyclic of order two.  Here $\TT\subset \CC$ is
the circle group of unit norm complex numbers.
  \end{example}

Let $\bP\to X$ be a principal $\bG$-bundle.  The obstruction to lifting to a
principal $G$-bundle is measured by a \emph{$\pi $-gerbe} $\sG(\bP)\to X$.
For $\Spin(n)\to SO(n)$ this $\pi $-gerbe is a geometric manifestation of the
second Stiefel-Whitney class.  There is a tautological construction of
$\sG(\bP)\to X$ as a sheaf.  (The reader may wish as a warmup to construct
directly from~$f$ the sheaf associated to the pullback $\pi $-bundle
in~\eqref{eq:7}.)  Let $U\subset X$ be an open set.  The value of~$\sG(\bP)$
on~$U$ is the collection of lifts of $(\bP\to X)\res U$ to a principal
$G$-bundle.  For small contractible~$U\subset X$ such lifts always exist, but
for general~$U$ there may be no lifts, in which case $\sG(\bP)(U)$~is empty.
More formally, an object in~ $\sG(\bP)(U)$ is a pair $(P_U\to U,\varphi )$
consisting of a principal $G$-bundle $P_U\to U$ and an isomorphism
  \begin{equation}\label{eq:17}
     \xymatrix{\bP\res U\ar[rr]^{\varphi }_{\cong }\ar[dr] && P_U/\pi \ar[dl]\\
     &U} 
  \end{equation}
We leave the reader to define the notion of an isomorphism $(P_U\to U,\varphi
)\to (P'_U\to U,\varphi ')$.  Thus $\sG(\bP)(U)$~is a groupoid and
$\sG(\bP)$~a sheaf of groupoids.  A global section is a lift of $\bP\to X$ to
a $G$-bundle.
 
Let $\Btp X$ denote the collection of $\pi $-gerbes over~$X$.  It is a
2-groupoid: there are isomorphisms of objects and isomorphisms of
isomorphisms.  For example, the groupoid of automorphisms of any $\pi
$-gerbe is the groupoid~$\Bp X$ of principal $\pi $-bundles over~$X$.  The
set of equivalence classes in~$\Btp X$ is the cohomology group~$H^2(X;\pi )$.
(Since $\pi $~is abelian, this cohomology group is well-defined.)  The group of
equivalence classes of automorphisms of any object is~$\Hp 1X$ and the group
of automorphisms of automorphisms is~$\Hp 0X$.  Homotopy groups are defined
for (higher) groupoids, and here
  \begin{equation}\label{eq:20}
     \begin{aligned} \pi _0\bigl(\Btp X \bigr)&\cong \Hp 2X \\ \pi
     _1\bigl(\Btp X\bigr) &\cong \Hp 1X 
      \\ \pi _2\bigl(\Btp X\bigr) &\cong \Hp 0X \\ \end{aligned} 
  \end{equation}

Let $\BNbG(X)$~denote the groupoid of $\bG$-connections on~$X$.  An object is
a principal $\bG$-bundle $\bP\to X$ with connection~$\bT\in \Omega
^1(\bP;\fg)$, which we simply denote as~$\bT$.  An isomorphism $\bT\to \bT'$
is an isomorphism $\varphi \:\bP\to \bP'$ of the underlying $\bG$-bundles
which satisfies~$\varphi ^*\bT'=\bT$.  The construction of the previous
paragraph defines a map 
  \begin{equation}\label{eq:18}
     \begin{aligned} p\:\BNbG(X)&\longrightarrow \Btp X \\ \bT&\longmapsto
      \bigl(\sG(\bT)\to X\bigr)\end{aligned} 
  \end{equation}
in which $\sG(\bT)$ denotes the $\pi $-gerbe associated to the $\bG$-bundle
carrying the connection~$\bT$. 

  \begin{definition}[]\label{thm:5}
 Fix $(\sG\to X)\in \Btp X$.  A \emph{relative field} over~$(\sG\to X)$ is a
pair~$(\bT,\theta )$ consisting of a $\bG$-connection~$\bT$ and an
isomorphism $\theta \:\sG\to \sG(\bT)$ of $\pi $-gerbes.  
  \end{definition}

\noindent
 The fiber of~$p$ over the trivial $\pi $-gerbe on~$X$ is the
groupoid~$B\mstrut _\nabla G(X)$ of $G$-connections on~$X$.

Relative fields form an ordinary groupoid---there are no automorphisms of
automorphisms.  We describe the automorphism group~$\rAut(\bT)$
of~$(\bT,\theta )$ in elementary terms.  As the notation suggests, this
automorphism group is independent of~$\theta $.  Suppose $\bP\to X$ is the
$\bG$-bundle which carries~$\bT$.  Conjugation $G\to\Aut(G)$ in~$G$ drops to
a group homomorphism $\bG\to\Aut(G)$, since $\pi \subset G$~is central.
There is an associated bundle of groups $\bP\times \mstrut _{\bG}G\to X$
associated to the conjugation action.  Sections of this bundle act on $\bP\to
X$, and $\rAut(\bT)$~is the stabilizer subgroup of~$\bT\in \Omega
^1(\bP;\fg)$.

  \begin{remark}[]\label{thm:6}
 Heuristically, the fields are $\bG$-connections with $G$-gauge
transformations.  Definition~\ref{thm:5} gives a precise formulation in terms
of \emph{local} fields.  We do not know a precise formulation in terms of
absolute fields.
  \end{remark}

We turn to the quantum theories~$\alpha $ and~$F$ built from these fields.
Fix a dimension~$n$.  The theory~$\alpha $ is an $(n+1)$-dimensional
topological theory and $F$~is a relative $n$-dimensional theory of
Riemannian\footnote{There may be spin structures and the theory may be
conformal, depending on particulars.} manifolds.  The topological
theory~$\alpha $ is defined by a finite path integral over $\pi $-gerbes;
see~\cite{Q,T,FHLT} for general discussions of homotopy finite quantum
theories.  If $W$~is a closed $(n+1)$-manifold, then
  \begin{equation}\label{eq:19}
     \alpha (W) = \frac{\#\Hp 2X\cdot \#\Hp 0X}{\#\Hp 1X}. 
  \end{equation}
Of course, this partition function is ignored in the truncation~$\ta$, so too
in $F\:\bo\to\ta$.
 
Let $X$~be a closed $n$-manifold.  Then $\alpha (X)$~is the vector space of
complex-valued functions on~$\Btp X$.  These are invariant functions on the
collection of $\pi $-gerbes, so they factor down to functions on equivalence
classes~$\pi _0\bigl(\Btp X \bigr)$.  The relative theory~$F$ gives an
element of~$\alpha (X)$, so a function 
  \begin{equation}\label{eq:21}
     F(X)\:\Hp 2X\longrightarrow \CC. 
  \end{equation}
 
Let $Y$~be a closed $(n-1)$-manifold.  Then $\alpha (Y)$~is the linear
category of vector bundles (of infinite rank) over~$\Btp Y$.  The relative
theory~$F$ determines a particular vector bundle 
  \begin{equation}\label{eq:22}
     F(Y)\longrightarrow \Btp Y. 
  \end{equation}
A complex vector bundle over a 2-groupoid only senses~$\pi _0$ and~$\pi _1$,
not~$\pi _2$ (nor higher homotopy groups if they were present).  If we choose
a basepoint in each component, then the fibers at the basepoints are complex
representations of~$\pi _1$, so can be decomposed according to the
irreducible representations.  For any finite abelian group~$A$, let
  \begin{equation}\label{eq:23}
     A\dual = \Hom(A,\TT) 
  \end{equation}
denote the \emph{Pontrjagin dual group} of characters.  Then, after choosing
basepoints, \eqref{eq:22}~determines topological vector spaces~$F(Y;m,e)$ for 
  \begin{equation}\label{eq:24}
     m\in \Hp2Y,\qquad e\in \Hp1Y\dual. 
  \end{equation}
The class~$m$ is a \emph{discrete magnetic flux} and $e$~is a \emph{discrete
electric flux}~\cite{W3}.  Note that if $Y$~is oriented, then
Poincar\'e-Pontrjagin duality is an isomorphism
  \begin{equation}\label{eq:25}
     \Hp1Y\dual\xrightarrow{\;\;\cong \;\;}\Hpd{n-2}Y. 
  \end{equation}

As in~\S\ref{sec:3} the picture of~$F$ as a \emph{boundary} theory
for~$\alpha $ is manifest in terms of classical fields.  If $W$~is an
$(n+1)$-manifold with boundary, and $\sG\to W$ a $\pi $-gerbe---a field in
the bulk theory~$\alpha $---then the boundary field is a $G$-connection
twisted by the restriction of the $\pi $-gerbe to the boundary.  In other
words, it is exactly~$(\bT,\theta )$ in Definition~\ref{thm:5} relative to
$\sG\res{\partial W}\to\partial W$.  The relative theory on an
$n$-manifold~$X$ has a bulk field $\sG\to[0,1]\times X$, a boundary
field~$(\bT,\theta )$ on~$\{1\}\times X$, and no additional field
 at~$\{0\}\times X$. 

There are disorder operators in the gauge theory with gauge group~$G$
associated to classes in~$H^2(X;\pi )$.  Suppose such a class is represented
by a normally oriented submanifold~$D\subset X$ of codimension two, together
with a locally constant function $j\:D\to\pi $.  Then, roughly speaking, the
value of~$F(X)$ in~\eqref{eq:21} on the class represented by~$(D,j)$ is the
path integral over $G$-connections on~$X\setminus D$ with limiting holonomy
prescribed by~$j$, where the limiting holonomy is computed around an oriented
circle surrounding~$D$ in the normal space as the radius of the circle
shrinks to zero.  As in~\S\ref{sec:3} this field defined on the complement
of~$D$ does not contain all the information that the relative field defined
on all of~$X$ does.
 
As in~\S\ref{sec:3} we can recover absolute (not relative) gauge theories
with gauge group~$G/\pi '$ for subgroups $\pi '<\pi $ from the relative
theory~$F$.  For example, the partition function on a closed $n$-manifold~$X$
is  
  \begin{equation}\label{eq:28}
     f_{\pi '}(X) = \sum\limits_{m'\in H^2(X;\pi
     ')}\frac{\#H^0(X;\pi ')}{\#H^1(X;\pi ')}\;F(X;m');
  \end{equation}
analogous to~\eqref{eq:12}.  There is a formula similar to~\eqref{eq:13} for
the quantum topological vector space. 
 
We use the language of Appendix~\ref{sec:6} to describe the relative fields.
Thus \eqref{eq:18} ~is a map
  \begin{equation}\label{eq:29}
     p\:\BNbG\longrightarrow B^2\pi 
  \end{equation}
with fiber~$\BNG$.  It is the classifying map for the principal $B\pi
$-bundle $\BNG\to \BNbG$.  The inclusion $i\:\pi '\to\pi $ of a subgroup
induces a pullback diagram 
  \begin{equation}\label{eq:30}
     \xymatrix{ B\mstrut _{\nabla }(G/\pi ')\ar[r]^{} \ar[d]_{p'} &
     \BNbG\ar[d]^{p} \\ B^2\pi '\ar[r]^{B^2i} & B^2\pi } 
  \end{equation}
On a manifold~$X$ integration over the fibers of~$p'$ followed by integration
over~$B^2\pi '(X)$ is equivalent to integration over the groupoid~$B\mstrut
_{\nabla }(G/\pi ')(X)$ of $G/\pi '$-connections on~$X$.

Finally, we briefly comment on the case when both $\alpha $~and $F$~are
homotopy finite quantum theories.  In this paragraph each topological space
is finite in the sense that it has only finitely many nonzero homotopy groups
and each is a finite group.  Let $B$~be a connected example of such a space
and $\alpha $~the $(n+1)$-dimensional theory defined by counting maps
into~$B$ up to homotopy, as in~\eqref{eq:19}.  A fibration $E\xrightarrow{\pi
} B$ defines a relative $n$-dimensional theory~$F_\pi \:\bo\to\ta$ using
relative $\sigma $-model fields up to homotopy.  Equivalently, it defines a
relative theory $\tF_\pi \:\ta\to\bo$.  Given a second fibration
$E'\xrightarrow{\pi '}B$, the composition 
  \begin{equation}\label{eq:38}
     \tF_{\pi '}\circ F_\pi \:\bo\to\bo 
  \end{equation}
is an (absolute) $n$-dimensional theory.  It can be described as the $\sigma
$-model into the total space of the fiber product of~$\pi $ and~$\pi '$.  In
terms of boundary conditions, the fibrations give an extra boundary field in
the $(n+1)$-dimensional $\sigma $-model with field~$f\:W\to B$, namely a lift
$g\:\partial W\to E$ of the restriction of~$f$ to the boundary.  The fields
in the $n$-dimensional theory~\eqref{eq:38} on an $n$-manifold~$X$ are a map
$f\:[0,1]\to B$, a lift $g\:\{0\}\times X\to E$ of $f\res{\{0\}\times X}$,
and a lift $g'\:\{1\}\times X\to E'$ of $f\res{\{1\}\times X}$.  The
canonical choices $B\xrightarrow{\pi '=\id}B$ and $*\xrightarrow{\pi '}B$
correspond roughly to Neumann and Dirichlet boundary conditions, and give
the absolute $n$-dimensional $\sigma $-models into~$E$ and the homotopy fiber
of~$\pi $, respectively.  Here $*$~denotes the contractible space of paths
in~$B$ which begin at a fixed point.  There is also an interpretation in
terms of gauge theory with gauge group~$\Omega B$.  Then the first boundary
condition is no condition at all and the second fixes the gauge at the
boundary.

   \section{Expectations for \TX}\label{sec:5}
% lastsubsec@000

In this section we fit the expectations~\cite[\S4]{W2} for \TX---the
$(0,2)$-superconformal theory in six dimensions---into a relative quantum
field theory.

A reductive real Lie algebra~$\fg$ is a direct sum $\fg=\mathfrak{z}\oplus
\fg'$ of its center~$\mathfrak{z}$ and its semisimple subalgebra
$\fg'=[\fg,\fg]$.  A Cartan subalgebra~$\fh'\subset \fg'$ determines a coroot
lattice, which is a full sublattice $\Gamma '\subset \fh'$.  Any two Cartan
subalgebras are conjugate by an element of~$\fg'$; the conjugation preserves
the coroot lattices.  A real Lie algebra with an invariant inner product is
reductive.

  \begin{data}[]\label{thm:7}
 \ 

  \begin{enumerate}
 \item A real Lie algebra~$\fg$ with an invariant inner product~$\form$ such
that all coroots have square length~2

 \item A full lattice $\Gamma \supset \Gamma '$ in~$\fh=\mathfrak{z}\oplus
\fh'$ such that the inner product is integral and even on~$\Gamma $
  \end{enumerate}
  \end{data}

\noindent
 The condition on the inner product implies that the semisimple
subalgebra~$\fg'$ is a sum of simple Lie algebras of ADE~type, i.e.,
$\fg'$~is \emph{simply laced}.  The lattice~$\Gamma $ may be specified by
choosing any Cartan subalgebra~$\fh'\subset \fg'$.  Any other choice is
conjugate by some $\xi' \in \fg'$, and we then conjugate~$\Gamma $ (and~$\fh$)
by the same element~$\xi '$, viewed as an element of~$\fg$.  A special case
is $\Gamma =\Gz\oplus \Gamma '$ for a chosen full lattice $\Gz\subset
\mathfrak{z}$.  

Given Data~\ref{thm:7}, define~$\Lambda \supset \Gamma $ as the dual lattice
to~$\Gamma $ in~$\fh$, the subset of vectors~$\eta \in \fh$ such that
$\langle \eta ,\Gamma \rangle\subset \ZZ$.  The quotient~$\pi =\Lambda
/\Gamma $ is a finite abelian group equipped with a perfect pairing 
  \begin{equation}\label{eq:31}
     \pi \times \pi \longrightarrow \QQ/\ZZ\subset \TT 
  \end{equation}
induced from~$\form$.  The pairing induces an isomorphism $\pi \cong \pi
\dual$.  In other words, $\pi $~is \emph{Pontrjagin self-dual}.  There is
also a quadratic form refining~\eqref{eq:31} induced from~$\langle -,-
\rangle$ which plays a role.  The data determine a covering~$G\to\bG$ of
compact connected Lie groups with kernel~$\pi $.  The lattice~$\Gamma $ is
the fundamental group of a maximal torus of~$G$ and $\Lambda $~is the
fundamental group of a maximal torus of~$\bG$.

There are two extreme cases worth noting.  If $\fg=\mathfrak{z}$~is abelian,
then Data~\ref{thm:7} reduces to a choice of lattice~$\Gz$ with a positive
definite even\footnote{There are 3-dimensional torus Chern-Simons theories
for which form~$\langle -,- \rangle$ is not even; they are defined on spin
manifolds.  We similarly expect that the evenness in Data~\ref{thm:7}(ii) can
be omitted to define a larger class of theories.} integral form~$\form$.  In
that case $G=\bG$~is the torus~$\mathfrak{z}/\Gz$.  The resulting
theory~$\sX_{(\Gz,\form)}$ is meant to be noninteracting.  At the other
extreme, if $\fg=\fg'$ is semisimple, so a sum of ADE~Lie algebras, then
$G$~is the 1-connected compact Lie group with Lie algebra~$\fg$, the finite
group~$\pi \in G$ is its center, and $\bG$~is the adjoint group.  We remark
that Data~\ref{thm:7} allows many intermediate cases, for example a lattice
in~$\mathfrak{s}\mathfrak{o}(4)$ for which $G=\bG=SO(4)$.

Now we can state the expected formal structure of \TX.

  \begin{expectation}[]\label{thm:8}
 Given~$(\fg,\form,\Gamma)$ in Data~\ref{thm:7} there exists a finite
7-dimensional topological quantum field theory $\ag=\alpha
_{(\fg,\form,\Gamma)}$ and a 6-dimensional quantum field theory
$\Xg=\sX_{(\fg,\form,\Gamma)}$ relative to~$\ag$. 
  \end{expectation}

\noindent 
 The topological theory~$\ag$ should be defined as an extended theory on a
bordism multi-category of manifolds which have a tangential structure to be
determined, roughly some sort of framing.\footnote{It will certainly include
an orientation and, at least in many cases, an integral Wu
structure~\cite{HS}.}  It is a 7-dimensional analog of 3-dimensional
Chern-Simons theory for torus groups~\cite{BM,FHLT,St}.  The 6-dimensional
theory~$\Xg$ is meant to be defined on a geometric bordism category of
manifolds with a conformal structure as well as some (topological) tangential
structure which includes a spin structure, since the theory has spinor
fields.  There is a conformal anomaly, and as for the chiral WZW theory
mentioned in~\S\ref{sec:2}, we treat it as a theory of Riemannian manifolds
and ignore the conformal anomaly. 
 
A useful formal picture is to imagine $\Xg$ as constructed from classical
relative fields\footnote{$\Xg$~has many other fields, which are well-defined.
We focus here on the self-dual 2-form field.}
  \begin{equation}\label{eq:32}
     p\:``{B^2_{\nabla }\bG}\textnormal{\,''}\longrightarrow B^3\pi , 
  \end{equation}
analogous to the quantization of relative $\sigma $-models and relative gauge
fields in the previous sections.  But while~$B^3\pi $ exists---a field $X\to
B^3\pi $ is a 2-gerbe with structure group~$\pi $ over~$X$---we do not know
an applicable notion of gerbes with connection to define~$B^2_\nabla \bG$
unless the Lie algebra~$\fg$ is abelian.  So \eqref{eq:32}~is in general only
a heuristic, whence the quotation marks.  

An important point is that both fields in~\eqref{eq:32} are to be treated as
\emph{self-dual}.  This is what restricts the dimensions of~$\Xg$ and~$\ag$
to~6 and~7, respectively.  The quantization procedure for self-dual fields is
not the standard Feynman functional integral, but rather involves a sort of
square root.  This is a well-defined and interesting story for abelian
fields, such as\footnote{The Pontrjagin self-duality~\eqref{eq:31} of~$\pi $
is needed to treat~$B^3\pi $ as a self-dual field.}~$B^3\pi $, and we hope to
develop it more elsewhere.  Here we indicate what such a development will
give for the finite topological theory~$\ag$.
 
Let $X$~be a closed oriented 6-manifold.  The cup product and
pairing~\eqref{eq:31} combine to a nondegenerate skew-symmetric pairing 
  \begin{equation}\label{eq:33}
     \Hp3X\times \Hp3X\longrightarrow \TT
  \end{equation}
on the finite group~$\Hp3X$.  This determines, up to noncanonical
isomorphism,\footnote{One role for the tangential structure in~$\ag$ will be
to make this construction canonical so that diffeomorphisms of~$X$ act
linearly, not just projectively.} a Heisenberg representation of a Heisenberg
central extension of~$\Hp3X$.  The finite dimensional vector space~$\ag(X)$
is the underlying vector space of this representation.
 
Let $Y$~be a closed oriented 5-manifold.  Then $\Hp3Y$~is the group of
self-dual fluxes.  To that end, notice from~\eqref{eq:31} and
Poincar\'e-Pontrjagin duality the isomorphisms 
  \begin{equation}\label{eq:34}
     \Hp3Y\cong \Hpd3Y\cong \Hp2Y\dual.
  \end{equation}
The first group may be thought of a magnetic and the last as electric, as
in~\eqref{eq:24}; the self-duality identifies them.  The linear
category~$\ag(Y)$ is the free $\Vi$-module with basis~$\Hp3Y$, and for each
self-dual flux $\sigma \in \Hp3Y$ we expect a quantum topological vector
space~$\Xg(Y;\sigma )$.

We turn now to dimensional reductions of \TX.

  \begin{claim}[]\label{thm:9}
 The dimensional reduction of~$\Xg$ to five dimensions is a relative gauge
theory based on $G\to\bG$ with kernel~$\pi $. 
  \end{claim}
 
\noindent 
 The particular gauge theory we obtain has maximal supersymmetry.  Our focus
here is on the assertion that it is a \emph{relative} quantum field theory,
so on the statement that the dimensional reduction of~$\ag$ is the finite
$\pi $-gerbe theory~$\alpha $ (in five dimensions) discussed
in~\S\ref{sec:4}.  To see this, observe that if $X=\cir\times \bX$ for a
closed oriented 5-manifold~$\bX$, then
  \begin{equation}\label{eq:35}
     \Hp3X\cong \Hp2{\bX}\times \Hp3\bX
  \end{equation}
has a canonical polarization, as written.  We can realize the Heisenberg
representation as functions on~$\Hp2\bX$, so identify~$\Xg(\cir\times \bX)$
as a function on~$\Hp2\bX$, which matches ~\eqref{eq:21}.  Similarly, if
$Y=\cir\times \bY$ is a closed oriented 5-manifold, then 
  \begin{equation}\label{eq:36}
     \Hp3Y\cong \Hp2\bY\times \Hp3\bY\cong \Hp2\bY\times \Hp1\bY\dual. 
  \end{equation}
We therefore obtain a quantum topological vector space~$\Xg(\bY;m,e)$ for
each $\sigma =(m,e)\in \Hp2\bY\times \Hp1\bY\dual$, which is consistent
with~\eqref{eq:24}. 

We conclude with three brief remarks.  The first is in~\cite[\S4.1]{W1}.  The
latter two explain why the two ways we can imagine extracting an absolute
quantum field theory from the relative theory~$\Xg$ fail.

  \begin{remark}[Reduction to four dimensions]\label{thm:10}
 Compactify on~$\cir\times \cir$ to obtain a four-dimensional relative gauge
theory.  For a closed oriented 4-manifold there are natural polarizations
of~$H^3(\cir\times \cir\times \bX)$ corresponding to primitive elements
of~$H^1(\cir\times \cir;\ZZ)$.  The obvious basis elements are exchanged
by~$\left(\begin{smallmatrix} 0&-1\\1&0 \end{smallmatrix}\right)$ acting on
the torus, which is implemented on the Heisenberg representation as Fourier
transform.  Using~\eqref{eq:28} to obtain ordinary gauge theories for
groups~$G/\pi '$, this Fourier transform exchanges~$\pi '$ and its orthogonal
complement~$\pi '^{\perp}$ relative to the perfect pairing~\eqref{eq:31}.
The groups~$G/\pi '$ and~$G/\pi '^\perp$ are Langlands dual. 
  \end{remark}

  \begin{remark}[]\label{thm:11}
 We passed from relative $\sigma $-models and relative gauge theories to
absolute theories by summing over a subgroup~$\pi '<\pi $.  The analog in the
self-dual situation involves a \emph{maximal} isotropic subgroup ($\pi
'^\perp=\pi '$), and these generally do not exist.
  \end{remark}

  \begin{remark}[]\label{thm:12}
 In two dimensions one tensors chiral and anti-chiral relative conformal field
theories to obtain an absolute conformal field theory.  This works with
supersymmetry since the vector representation is reducible in two dimensions.
Thus the tensor product of a theory with $(p,0)$-supersymmetry and one with
$(0,q)$-supersymmetry has $(p,q)$-supersymmetry.  But in six dimensions the
vector representation is irreducible, and the tensor product of theories with
$(p,0)$-supersymmetry and $(0,q)$-supersymmetry has a symmetry group which
contains two copies of the ordinary Poincar\'e group.   The diagonal subgroup
does not admit an extension to a super Poincar\'e group which is a symmetry
of the theory.  In short, the tensor product has no supersymmetry.
  \end{remark}

\appendix

   \section{What is a classical field?}\label{sec:6}
% lastsubsec@000

A ``scalar field'' or ``gauge field'' is not specific to a particular
manifold, but rather is defined for all manifolds~$X$.  Furthermore, fields
pull back under maps $X'\to X$ and they are \emph{local}: compatible fields
on open sets glue uniquely into a field on the union.  The geometric object
which encodes this locality is a \emph{sheaf}.  However, it is not a sheaf
defined only on open subsets of a fixed space, but rather a sheaf~$\sF$ which
one evaluates on smooth manifolds.  The value~$\sF(X)$ on a smooth
manifold~$X$ is a \emph{set} for a scalar field---the set of real-valued
functions on~$X$---but for a gauge field $\sF(X)$~is the \emph{groupoid} of
connections on~$X$.  A leisurely introduction to sheaves in this context may
be found in~\cite{FH}.

Some fields---scalar fields, gauge fields---pull back under arbitrary smooth
maps of manifolds, and the manifolds can have any dimension.  Others, such as
metrics, require both that the manifolds be of the same dimension and that
the map be a local diffeomorphism.\footnote{Metrics also pull back under
immersions.  But we treat tangential structures, such as orientations, as
fields.  After all, they too pull back and satisfy a gluing condition.
Orientations of a manifold only pull back under maps of manifolds of the same
dimension.}  Therefore, we are led to the following definition of the domain
category for classical fields in an $n$-dimensional field theory.

  \begin{definition}[]\label{thm:14}
  For each integer~$n\ge 0$ define~$\Man_n$ as the category whose objects are
smooth $n$-manifolds and morphisms are local diffeomorphisms. 
  \end{definition}

We have already mentioned that $G$-connections on a manifold~$X$ form a
groupoid~$\BNG(X)$.  In this paper we encounter fields which form a higher
groupoid: $B^2\pi (X)$~in~\S\ref{sec:4} is a 2-groupoid and $B^3\pi
(X)$~in~\S\ref{sec:5} is a 3-groupoid.  It is convenient to view all of these
as having values in the category~$\sSet$ of simplicial sets.  For the
purposes of this paper the reader need only be aware that we need some
mathematical object which tracks internal symmetries of fields, symmetries of
symmetries, etc. 
 
This discussion is summarized in a succinct statement.

  \begin{definition}[]\label{thm:13}
 A classical field, or collection of classical fields, in an $n$-dimensional
field theory is a simplicial sheaf~$\sF\:\Man_n\op\to\sSet$.
  \end{definition}

\noindent
 Any homomorphism $\sF\:\Man_n\op\to\sSet$ is called a \emph{presheaf}; the
sheaf condition, which we do not spell out here (see~\cite[\S5]{FH}),
expresses the locality with respect to gluing on open covers.  As mentioned,
some fields---including all those encountered in this paper---extend to a
sheaf on the category of all smooth manifolds and smooth maps.  A map of
fields, such as~\eqref{eq:15} and~\eqref{eq:29}, is defined as a natural
transformation of functors.

%\appendix

\providecommand{\bysame}{\leavevmode\hbox to3em{\hrulefill}\thinspace}
\providecommand{\MR}{\relax\ifhmode\unskip\space\fi MR }
\providecommand{\MRhref}[2]{%
  \href{http://www.ams.org/mathscinet-getitem?mr=#1}{#2}
}
\providecommand{\href}[2]{#2}

  \end{document}